# Computationally efficient spatial rendering of late reverberation in virtual acoustic environments

Christoph Kirsch
*Medizinische Physik
and Cluster of Excellence Hearing4all,
Carl von Ossietzky Universität*
Oldenburg, Germany
christoph.kirsch@uol.de

Josef Poppitz
*Akustik
and Cluster of Excellence Hearing4all,
Carl von Ossietzky Universität*
Oldenburg, Germany
poppitz@itap.de

Torben Wendt
*Medizinische Physik and Akustik
and Cluster of Excellence Hearing4all,
Carl von Ossietzky Universität*
Oldenburg, Germany
torben.wendt@uol.de

Steven van de Par
*Akustik
and Cluster of Excellence Hearing4all,
Carl von Ossietzky Universität*
Oldenburg, Germany
steven.van.de.par@uol.de

Stephan D. Ewert
*Medizinische Physik
and Cluster of Excellence Hearing4all,
Carl von Ossietzky Universität*
Oldenburg, Germany
stephan.ewert@uol.de

*Abstract*— For 6-DOF (degrees of freedom) interactive virtual acoustic environments (VAEs), the spatial rendering of diffuse late reverberation in addition to early (specular) reflections is important. In the interest of computational efficiency, the acoustic simulation of the late reverberation can be simplified by using a limited number of spatially distributed virtual reverb sources (VRS) each radiating incoherent signals. A sufficient number of VRS is needed to approximate spatially anisotropic late reverberation, e.g., in a room with inhomogeneous distribution of absorption at the boundaries. Here, a highly efficient and perceptually plausible method to generate and spatially render late reverberation is suggested, extending the room acoustics simulator RAZR [Wendt et al., J. Audio Eng. Soc., 62, 11 (2014)]. The room dimensions and frequency-dependent absorption coefficients at the wall boundaries are used to determine the parameters of a physically-based feedback delay network (FDN) to generate the incoherent VRS signals. The VRS are spatially distributed around the listener with weighting factors representing the spatially subsampled distribution of absorption coefficients on the wall boundaries. The minimum number of VRS required to be perceptually distinguishable from the maximum (reference) number of 96 VRS was assessed in a listening test conducted with a spherical loudspeaker array within an anechoic room. For the resulting low numbers of VRS suited for spatial rendering, optimal physically-based parameter choices for the FDN are discussed.

*Keywords—room acoustics simulation, psychoacoustics, complex acoustic environments, loudspeaker array, interaural coherence*

## I. Introduction

Room acoustics simulation and virtual acoustic environments (VAEs) enable the auralization of existing or not (yet) existing spaces with applications ranging from architectural planning [1] to entertainment [2]. Because of their potentially high ecological validity [3]–[5], VAEs have gained interest as tools for psychoacoustic research and hearing aid development [6]–[9].

For many of these applications, computational efficiency is important to allow for interactive real-time updates with 6-DOF (degrees of freedom) movement of sources and receivers. Thus, simplifications of the underlying room acoustics simulation and the rendering are desirable.

In enclosed spaces, due to reflections at boundaries, besides direct sound, early reflections and reverberation occur. Whereas individual early reflections might be directional in nature, late reverberation results from a superposition of many densely spaced reflections that are spatially more or less evenly distributed. Depending on room geometry and the spatial distribution of sound absorption at the boundaries, the resulting diffuse late reverberation can be considered spherically isotropic or might contain limited spatial directivity [10]–[13]. Thus, for room acoustics simulation and rendering of sound fields in VAEs, the spatial resolution required for rendering diffuse late reverberation is of interest. Applications focusing on human perception and behavior in the VAE might allow for a reduction of spatial resolution for late reverberation without altering the perceived sound field, enabling simplifications of the simulation and rendering.

Room acoustics simulations are often based on geometric acoustics in the form of image source models (ISM) [14], [15], ray tracing [16], [17], radiosity [18] and radiance transfer [19]. Based on statistical assumptions, computationally highly efficient reverberation algorithms like feedback delay networks (FDN; [20]) can be applied for generating reverberated signals (e.g., [21], [22]). In hybrid approaches, simplified statistical methods, which fulfill perceptual requirements for late reverberation can be combined with more accurate geometric acoustic models for early reflections. An ISM for early reflections and raytracing to calculate the spatiotemporal energy distribution of scattered reflections and late reverberation were combined in [23]. An ISM and a spatialized FDN for late reverberation were combined in [24]. Perceptual evaluations in [25] showed that a high degree of perceptual plausibility can be achieved with current room acoustics simulation methods.

The spatial resolution of late reverberation simulation and rendering is determined by the number of (incoherent) late reverberation signals which are spatially mapped around the listener. Such virtual reverberation sources (VRS) can be played back by individual loudspeakers in an array. For VRS at positions differing from those of physically available loudspeakers in a given array, VRS can be rendered to multiple loudspeakers using vector-base amplitude panning (VBAP; [26]), or spherical-harmonics-based approaches (Ambisonics; [27], [28]). Using these methods, a diffuse sound field can be approximated by superposition of incoherent sounds from many directions (for a review of

This work was funded by the Deutsche Forschungsgemeinschaft, DFG – Project-ID 352015383 – SFB 1330 C5.

theoretical approaches see, e.g., [29]). Regarding the perception of diffuse sound fields, [30] found that a specific horizontal arrangement of only four loudspeakers separated by 90° can already be sufficient to reproduce the spatial impression of a diffuse sound field. With such a low number of loudspeakers, however, the results were strongly dependent on the rotation of the loudspeaker array with regard to the listener. This is problematic in VAE applications, where listeners can freely rotate their head in the array.

In the room acoustics simulator (RAZR; [24]), incoherent late reverberation signals are generated as output of an FDN and used to drive a limited number of VRS which are spatially evenly distributed on a listener-centred cube. Alternatively, other approaches use, e.g., incoherent decaying noises with a spectral profile matching the frequency-dependent reverberation time $RT_{60}$ as filters to drive a limited number of VRS. While in the case of a completely homogeneous distribution of absorption coefficients at all boundaries an isotropic late reverberant field has to be approximated with a limited number of VRS, the more critical case regarding spatial resolution of late reverberation, however, is an inhomogeneous distribution of absorption coefficients, like an opening at one side of a room or one highly absorbing wall.

In the current study, a computationally highly efficient method to generate and spatially render isotropic and ansiotropic late reverberation is suggested. For simplicity a "proxy" shoebox approximation of room geometry is used in the context of the room acoustics simulator RAZR. The spatial resolution required for rendering late reverberation was assessed in a simulated room with one highly absorbing wall covering different solid angles or field of view (FOV) from the listener position. The number of VRS was varied from low spatial resolution (6 VRS) to high spatial resolution (96 VRS). A simplistic, highly efficient spatial subsampling method was suggested to map the spatial distribution of reflection coefficients at the room boundaries to weighting factors for the VRS. A 3-dimensional, 86-channel loudspeaker array was used for rendering. The VRS as well as direct sound and early specular reflections up to the third reflection order were mapped to the loudspeaker array using VBAP. For technical evaluation, the coherence between the two ears of a dummy head in the sound field generated by the loudspeaker array was analyzed. Interaural coherence is considered relevant for psychoacoustic processes (e.g., [31], [32]) and also has been suggested to assess the reproduction of diffuse sound fields (e.g., [30], [33]). In addition, the frequency-dependent interaural level difference (ILD) was assessed. A psychoacoustic experiment with normal-hearing listeners was conducted to investigate whether listeners can discriminate renderings with a lower number of VRS from the highest available number of 96 VRS, which served as a reference.

II. ROOM ACOUSTICS SIMULATION

*A. Basic method*

RAZR [24] generates perceptually plausible room impulse responses (RIRs) using a hybrid ISM/FDN approach. To achieve high computational efficiency, the ISM uses a shoebox approximation of room geometry. By default, RAZR computes discrete early (specular) reflections up to the third reflection order in the ISM. The reverberation tail is generated by a 12-channel FDN which is fed by the last order of reflections from the ISM. The output channels of the FDN are used as VRS that are spatially mapped to the same number of discrete locations on a listener centered cube, aligned with the six walls of the "proxy" shoebox room (reverberation cube mapping). In this default setting, the spatial resolution of the late reverberation is a result of 12 discrete directions (2 representing each wall) spatially evenly distributed around the listener. The frequency-dependent absorption coefficients of each of the 6 walls determines a wall-specifc "absorption filter" which is applied to the respective 2 FDN outputs in the 2 VRS representing each wall. Perceptual plausibility of the resulting room acoustics simulations was demonstrated by favorable performance in comparison to other state-of-the-art approaches in ([25], see their Fig. 8).

*B. Suggested extensions*

To assess the required spatial resolution of late reverberation rendering, in the current study RAZR was adapted to generate either 6, 12, 24, 48, or 96 VRS, equivalent to 1, 2, 4, 8, and 16 VRS per surface of the underlying shoebox room. To ensure similar spectro-temporal characteristics of the FDN output throughout evaluation and avoiding timbre changes in the resulting reverberant tail due to a variation of the number of FDN channels ([34], [35]), the FDN always operated with the highest number of 96 channels, independent of the number of rendered VRS. The 96 FDN outputs were directly used for 96 VRS and the spatial resolution of the late reverberation was adjusted by using a set of downmixes for 48, 24, 12, and 6 VRS generated in a sequential procedure by adding pairs of the FDN output channels. The VRS were spatially distributed according to the directions of vertices of polyhedra, centred on the listener and directionally aligned with the (shoebox) room boundaries. The polyhedra had a number of vertices equivalent to the number of VRS and were optimized for sphericity [36], which ranged from 0.86 for 6 VRS to 0.99 for 96 VRS. For 6 VRS, the resulting directions are orthogonal to each other and for 12 VRS they correspond to points lying on the diagonals of the surfaces of a room aligned cube (similar to the original alignment suggested in [24]). Directions for 24 and more VRS were based on a combination of 1, 2, and 4 snub cuboctahedra (snub cubes).

In rooms with an inhomogeneous distribution of acoustic absorption, e.g., a shoebox room with one highly absorbent wall and otherwise more reflective walls, anisotropic late reverberation will occur. The spatial distribution of late reverberation depends on the distance of the listener to the absorbing wall, affecting the solid angle (or field of view, FOV) occupied by the wall. In the extreme case of a listener very close to one completely absorbing wall, no reverberant sound energy impinges from the hemisphere occupied by the absorbing wall. To cope with the general case of different arbitrary absorption coefficients per wall (representing the average absorption coefficient of the room geometry approximated by the walls of the proxy shoebox room), RAZR was extended by a spatial sampling routine to calculate the absorption filters applied to each of the VRS. Dependent on the listener's position in relation to the boundaries, the resulting FOV occupied by each boundary and the direction of the VRS in relation to the boundaries, the contribution of the absorption coefficients to the absorption filter at the FDN output for the respective VRS were blended.

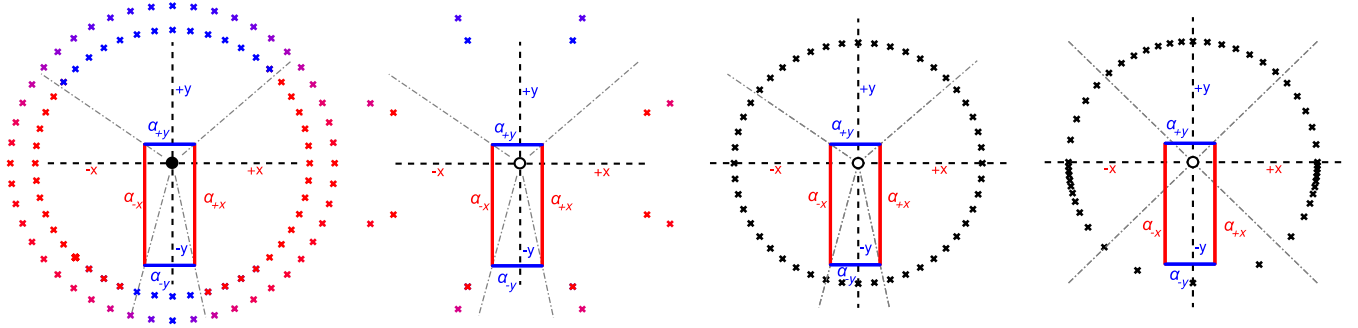

Fig. 1. Left panels: Distribution of VRS (crosses) around the listener (black circle) in a room (rectangle). Colors represent reflection coefficients associated with particular boundary surfaces of the room. Inner circle: simple assignment of VRS to a certain boundary based on projection. Outer circle: smooth weighted assignment based on warping on the VRS directions. Right panels: VRS positions (black crosses) and directions of the room corners (dash-dotted lines) before warping (mid-right) and after warping (right-most).

Fig. 1 shows a 2-dimensional representation of an example case for a number of 48 VRS (left-most panel) and 8 VRS (mid-left panel) in the horizontal plane. Here, an elongated room is shown and the receiver is placed close to the upper $+y$ wall. Red and blue indicate the vertical and horizontal walls, respectively, as well as their associated attributes. The black dashed lines are along the x and y directions and coincide with the (axis-aligned) wall normals. The inner circle of red and blue colored crosses shows a straight-forward approach where the absorption coefficients of a respective walls are directly projected onto the VRS. While this simplistic approach appears feasible for a high number of VRS (left-most panel), it is unsuited for lower numbers of VRS (mid-left panel), where the lower $-y$ wall would not be represented. Thus, an efficient spatial sub-sampling has to be devised. Additionally, smooth transitions are desired for dynamic environments, in which the receiver can move. To achieve a smooth sampling for arbitrary number and positions of VRS, the space around the receiver position was divided in octants (quadrants in the depicted 2-dimensional case in Fig. 1, as indicated by the dashed lines). In each octant, the VRS direction vectors (from receiver to VRS) were divided elementwise by the vector from the receiver to the shoebox corner (vertex) contained in the octant, and then normalized. The resulting warped VRS position of index k can be expressed as

$$\overrightarrow{VRS}_{w,k} = \frac{\overrightarrow{VRS}_k \oslash \vec{V}_{n,abs}}{\|\overrightarrow{VRS}_k \oslash \vec{V}_{n,abs}\|}, \quad (1)$$

where $\overrightarrow{VRS}_k = [x_k\ y_k\ z_k]$ is the original VRS position, $\oslash$ is the Hadamard (elementwise) division operator, and $\vec{V}_{n,abs}$ is the vector of absolute values of the room vertex coordinates relative to the receiver in the octant n. This transformation warps the vectors from the receiver to the eight vertices of the shoebox to represent a receiver-centred cube. Likewise, all k VRS directions are warped in such a way that a VRS on the room corner appears exactly on the according cube vertex. The two panels on the right-hand side of Fig. 1 (mid-right: before warping, right-most: after warping) show the effect of warping in the depicted 2-dimensional representation. The dash-dotted lines to the edges (vertices) appear under an angle of 45° exactly on the diagonals between the normals of the two walls in each quadrant after warping (right-most panel).

Using the warped VRS directions, the contributions of the reflection coefficients of each of the orthogonal shoebox walls are calculated using VBAP with the inverted (outside-pointing) wall normals as vector base. Eq. 18 from [26] is therefore adapted to yield the contribution $g_k = [g_1\ g_2\ g_3]$ of the absorption coefficients of the three intersecting walls in each octant with the normal vectors $n_1, n_2, n_3$ to the filtering applied to a particular VRS with index k:

$$g_k = \overrightarrow{VRS}_{w,k} \begin{bmatrix} n_{1,x} & n_{1,y} & n_{1,z} \\ n_{2,x} & n_{2,y} & n_{2,z} \\ n_{3,x} & n_{3,y} & n_{3,z} \end{bmatrix}^{-1}. \quad (2)$$

For an axis-aligned shoebox room, $g_k$ simplifies to the absolute values of each element of $\overrightarrow{VRS}_{w,k}$. The resulting spatially (sub-) sampled absorption coefficient $a_k$ (typically computed for multiple frequencies) which serves as a basis for designing the absorption filter for each VRS is

$$a_k = g_k \begin{bmatrix} \alpha_1 \\ \alpha_2 \\ \alpha_3 \end{bmatrix}, \quad (3)$$

where $\alpha_1, \alpha_2, \alpha_3$ are the absorption coefficients of the respective walls in the octant. Following this method, a VRS at a room corner receives an identical contribution from all three walls (two in the depicted 2-dimensional case). In a case where the listener is close to a highly absorbent boundary, the corresponding VRS in the FOV, defined by the "view" frustum from the receiver to the four shoebox vertices of the boundary now emits significantly less energy than the other VRS, mimicking that no or reduced sound energy in the late reverberant field can emerge from that boundary. The further the listener moves away from the absorbent boundary, the smaller the surface's FOV and the fewer VRS are affected. Depending on the overall number of VRS, the spatial resolution with which the absorbing boundary, as observed from the listener position, is spatially sampled by the VRS directions is reduced. This results in an increasingly blurred spatial representation with decreasing number of VRS.

It is assumed that the highest number of 96 VRS applied here is sufficiently high to serve as a reference condition for evaluation of spatial resolution of late reverberation with all 96 FDN channels mapped to separate VRS.

III. EVALUATION SETUP

A. Virtual Room

A corridor (Fig. 2) with the dimensions 24 m x 8 m x 6 m and inhomogeneous absorption coefficients was used for the evaluation of the proposed method. One of the small surfaces (shaded) at the end of the corridor was highly absorbent ($\alpha$ = 0.99 for all frequencies), while all other surfaces were quite reflective ($\alpha$ = 0.01 to 0.11 from 125 Hz to 8 kHz). The resulting $RT_{60}$ ranged from approx. 1.3 s to 0.8 s in the frequency range from 125 Hz to 8 kHz. Four source-receiver-position combinations were considered to vary the solid angle (or FOV) occupied by the highly absorbing wall and thus the spatial features of the resulting anisotropic late reverberation. Source and receiver were always aligned on an axis parallel to the highly absorbent surface, and were located at a height of 1.8 m above the floor and 1.33 m from the sidewalls, resulting in a fixed source-receiver distance of 5.33 m. There were four source-receiver combinations at different distances to the highly absorbing wall, so that a wide range of different FOVs of the absorbent wall is obtained. The default receiver orientation was towards the sound source (azimuth angle of 0°) for all combinations (denoted A to D, see also Fig. 2).

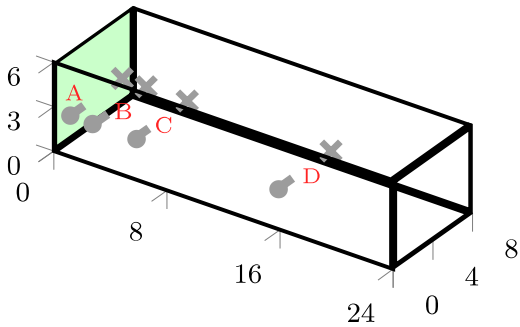

Fig. 2. Virtual Corridor. Source positions are denoted by crosses, receiver positions and orientation by the nosed balls. The highly absorbing surface is shaded. The distances to highly absorbing surface were chosen so that the horizontal FOV was approximately 170°, 110°, 70° and 30° for position A, B, C, and D, respectively.

### B. Loudspeaker Array Rendering

The room acoustic simulations of the corridor, composed of the discrete early reflections and the (varying number) of VRS were rendered using VBAP to a spherical 3-dimensional, 86-channel loudspeaker array (Genelec 8030 c/b), mounted in a 7 m x 9 m x 7 m anechoic chamber with 0.75 m foam wedges. The radius of the array is 2.5 m. The loudspeakers are inhomogeneously arranged in five rings at -60°, -30°, 0°, 30°, 60° elevation and two additional loudspeakers below and above the center point (-90°, 90° elevation). The azimuthal spacing of loudspeakers is 7.5° in the horizontal ring and 30° and 60° respectively in the rings outside of the horizontal plane. The simulated RIRs that have been spatialized for the loudspeaker array are referred to as multichannel room impulse responses (MRIRs) in the following. Due to path differences between the loudspeakers contributing in the rendering of a VRS and the listener's ears, spectral coloration artefacts can occur when using VBAP. In RAZR, these are accounted for with filtering based on a statistical approach described in [37].

## IV. TECHNICAL EVALUATION

Interaural coherence and interaural level difference (ILD) were estimated using the FABIAN HRTF database [38] in the late reverberant sound field (disregarding early reflections and direct sound) rendered using the loudspeaker topology of the 86-channel loudspeaker array. For this technical evaluation, independent Gaussian noises were used as output of the VRS, which received the weights according to the suggested spatial sampling method.

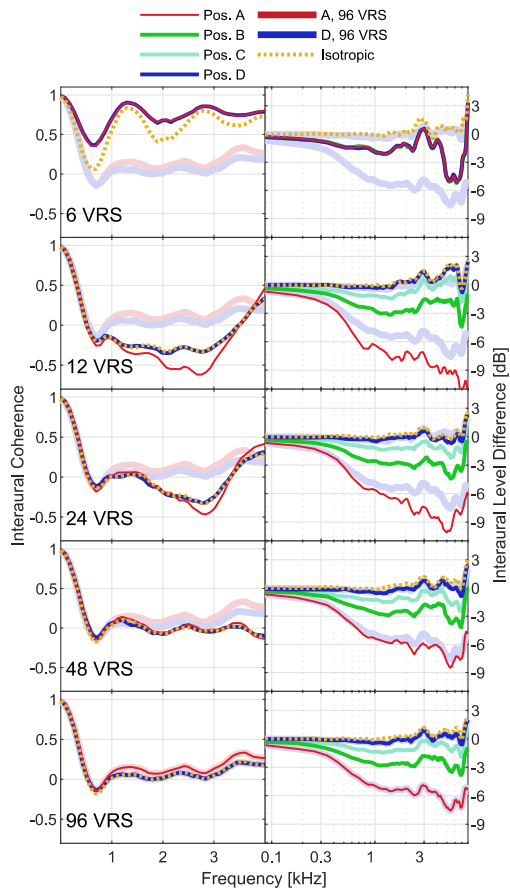

Fig. 3. Interaural coherence (left column) and interaural level difference (right column) for 6 to 96 VRS (top to bottom) and different positions in the corridor as indicated in the legend. The dotted yellow trace shows the approximation of an isotropic sound field with varying number of VRS in the current setup. For reference the results for 96 VRS for position A and D are displayed by less saturated thick lines in all panels.

The frequency-dependent interaural coherence (IC) estimates $C_{lr}(f)$ between the signals $l$ and $r$ in both ears were calculated according to:

$$C_{lr}(f) = \Re\left(\frac{G_{lr}(f)}{\sqrt{G_{ll}(f)G_{rr}(f)}}\right). \quad (4)$$

where f denotes the frequency, $\Re$ is the real-part operator and G represents the spectral density estimate according to Welch. The calculations were performed for consecutive windows with a 75 % overlap and a length of 512 samples at 44.1 kHz sampling rate to obtain an average coherence estimate. The Gaussian noise VRS signals were 60 s in duration.

The ILD was calculated from the same simulated ear signals $l$ and $r$ like above as

$$\text{ILD}(f) = 10\log_{10}{G_{ll}(f)}/{G_{rr}(f)}. \quad (5)$$

The left column of Fig. 3 shows the IC for increasing numbers of VRS (top to bottom) for position A (close to the absorbing wall) and position D (furthest away from the absorbing wall). Intermediate results are observed for positions B and C which are not shown for clarity. The dotted yellow trace represents the approximation of the isotropic case in the current rendering setup with the same output power assigned to all VRS instead of the suggested weighting according to the inhomogeneous distribution of absorption coefficients at the boundaries. The results for positions

Tab. 1. Heuristically determined frequency limit for reasonably accurate sound reproduction related to 96 VRS rendering. All values are in Hz.

|     | IC   |      | ILD  |        |
| --- | ---- | ---- | ---- | ------ |
| VRS | A    | D    | A    | D      |
| 6   | 400  | 400  | 400  | 400    |
| 12  | 950  | 950  | 600  | 5000   |
| 24  | 1800 | 1800 | 1500 | > 8000 |
| 48  | 3000 | 3000 | > 8000 | > 8000 |

A and D for 96 VRS are shown for comparison in all panels (less saturated thick traces). For 6 VRS (top panel), large deviations from the 96 VRS condition are obserKirsch et al.ved above about 400 Hz and no difference is obtained for positions A and D. For an increasing number of VRS, the IC gradually asymptotes against that obtained for 96 VRS with deviations only occurring at increasingly higher frequencies. Tab. 1 shows the upper frequency limit above which larger deviations occur. Differences between position A and D also become are increasingly like those observed for 96 VRS with increasing number of VRS. Overall, as shown for the "reference" case of 96 VRS, only small differences in IC for positions A and D are observed, although no sound energy impinges from one hemisphere in position A. In position D the results already are virtually identical with those observed for the isotropic case (dashed).

The right column of Fig. 3 shows the ILD for different numbers of VRS and all positions A to D. Generally, ILDs are largest for position A (close to the absorbent wall) and decrease to near zero for the farthest position D, similar to the ILD observed for the approximation of the isotropic case (dashed). Remaining small ILDs are to be attributed to the HRTF-database used for simulation and the selection of the nearest-neighbor directions in the database (according to the loudspeaker directions in the array). Again, large deviations from 96 VRS are mainly observed for 6 VRS where the maximum ILD for position D (up to 5 dB) in the depicted 8 kHz range is considerably smaller than for 96 VRS (up to 8 dB). For 12 VRS, a larger maximum ILD compared to 96 VRS (of up to 10 dB) is observed, and the results for larger numbers of VRS gradually asymptote against those observed for 96 VRS. The right-hand columns in Tab. 1 indicate the upper frequency limit of reasonable agreement with the 96 VRS case for A and D, which increases for increasing number of VRS.

In Fig. 4, IC (left column) and ILD (right column) are displayed for different receiver orientations to assess the effect of rotational movements in the VAE. The orientation of 0° (replot of traces from Fig. 3) refers to the orientation indicated in Fig. 2. Orientations of 30° (dashed) and 60° (dotted) are clockwise rotations towards the corridor. It is obvious that IC in the left column is independent of rotation up to a frequency limit which increases with the number of VRS. For 6 VRS, IC for different rotations starts diverging around 500 Hz, whereas this frequency is about 900 Hz for 12 VRS, 2000 Hz for 24 VRS, and 3000 Hz for 48 and 96 VRS. The ILDs depicted in the left column of Fig. 4 show that for any number higher than 6 VRS, there are no major deviations in ILD depending on orientation for the near-isotropic position D. An overall similar behavior is observed for all orientations for more than 6 VRS at position A. For 6 VRS, strong deviations from 96 VRS are observed for all orientations.

In summary, the technical evaluation shows a large deviation for 6 VRS from the higher numbers of VRS. The IC in the perceptually relevant range up to 1.5 kHz (e.g., [39]–

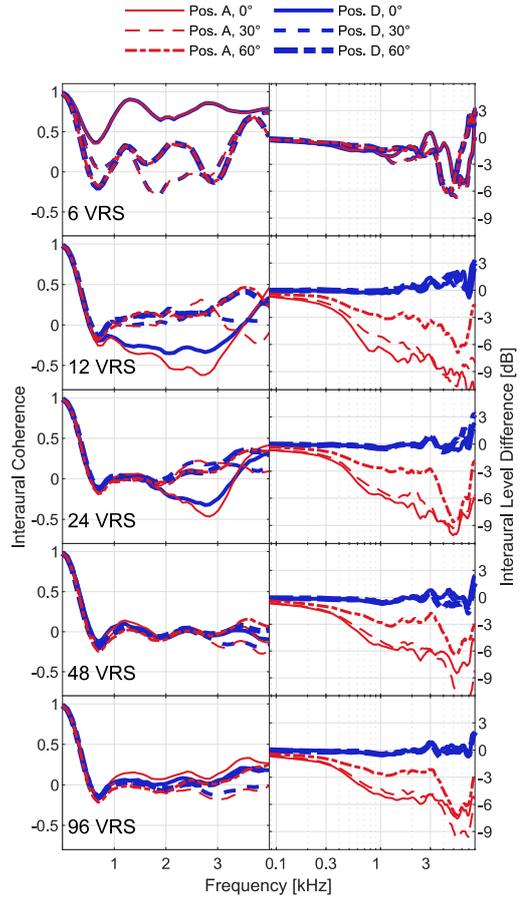

Fig. 4. Interaural coherence (left column) and interaural level difference (right column) for 6 to 96 VRS (top to bottom) for different receiver orientation. 0° refers to the orientation indicated in Fig. 2 (replot from Fig. 3). 30° and 60° are a clockwise rotation towards the corridor.

[41]) is well approximated for 24 VRS and more, while 12 VRS still show deviations above about 750 Hz (especially when receiver rotation is considered). However, these deviations mainly occur for small coherence values where perceptual discrimination is poor (e.g., [42]). Regarding ILDs, a good approximation is also obtained for 24 VRS and more. For 6 VRS, considerably smaller ILDs compared to 96 VRS are observed, while for 12 VRS larger ILDs are observed.

## V. PERCEPTUAL EVALUATION

### A. Method

*1) Listeners:* 6 listeners (3 female, 3 male) were recruited amongst students at the University of Oldenburg. They were between 21 and 28 years old. All were tested for normal hearing by means of pure tone audiometry (Hearing threshold $\leq 25$ dB HL between 125 Hz and 8 kHz). All subjects reported varying amounts of experience with listening tests.

*2) Stimuli:* MRIRs were generated for the four source-receiver conditions A-D (see Fig. 1) with varying numbers of 6, 12, 24, 48 and 96 VRS but unaltered early reflections for each of the positions. A transient stimulus referred to as pink pulse was chosen as a source signal for convolution with the MRIRs. The signal was generated as a digital delta pulse that was subsequently filtered in order to achieve a pink spectrum (sampling rate 44.1 kHz, decay from 0 dB FS to -60 dB FS in 36 ms). The pink pulse provides listeners with a highly consistent percept that due to its transient nature can be assumed to result in high sensitivity to alterations in the spatial rendering of the reverberant decay.

*3) Apparatus and Procedure:* Listeners were seated on a fixed (non-rotating) chair in the center of the above described 86-channel spherical loudspeaker array.

A computer monitor was placed straight ahead of the listeners in a distance of 2.5 m in order to inform them about the progress of the experiment and to provide a fixed direction for default head orientation and gaze prior to the presentation of the stimulus. The listeners' head movement was, however, neither constrained nor monitored, enabling natural head movements during listening. Listeners used a wireless keyboard on their lap so that responses could be provided without looking at the controls.

For the listening tests, an ABX paradigm was used. The reference rendering with the maximum number of 96 VRS and the rendering under test (with lower number of VRS) were presented randomly as either A and B of the sequence. X was randomly chosen to be either the reference or the rendering under test. Test subjects had to determine whether X was perceived to be similar to A or B in terms of spatial properties.

The procedure was separated into 4 runs that each covered a particular room condition. Listeners had the opportunity to take short breaks in between runs. 20 presentations per number of VRS resulted in 80 ABX trials per run. One experimental run took about 10 minutes. Listeners did not receive any feedback on the correctness of their responses. Prior to the listening test, audiometry and a familiarization phase were performed. The familiarization phase consisted of the presentation of four ABX trials with 6 and 12 VRS. During the familiarization phase, listeners received feedback on the correctness of their responses. The experimental procedure, except for the audiometry, was repeated in an additional session on a different day resulting in overall 40 presentations per VRS number.

*B. Results*

Fig. 4 shows the average discrimination results between conditions with fewer VRS as indicated on the x-axis and the reference condition with 96 VRS. The order of the boxes and colors indicate the room conditions with different distance from the absorbing wall (left to right: A-D, large to small FOV) as indicated in the legend, grouped by the number of VRS.

A fairly narrow distribution of the results can be observed with 6 VRS for all positions. For these stimuli, all listeners responded correctly in at least 95 % of the trials. For 12 VRS, there is a wider spread for all positions with individual performance ranging between 52.5 % and 87.5 % correct responses. A generally wide performance range can also be observed for 24 and 48 VRS, but with lower medians compared to 12 VRS. A consistent dependency on the position within the room (A-D) is not obvious.

A two-way repeated-measures was performed to assess the effect of the number of VRS and position in the room. A significant main effect was found for the number of VRS [$F(3, 15) = 288.87$, $p < 0.001$]. No significant main effect of the position [$F(3, 15) = 0.64$, $p = 0.36$] and no significant interaction [$F(9, 45) = 1.18$, $p = 0.49$] were found. Post-hoc pair-wise comparisons (Bonferroni) revealed significant differences between the conditions with 6 VRS compared to all other conditions as well as between 12 VRS and all other

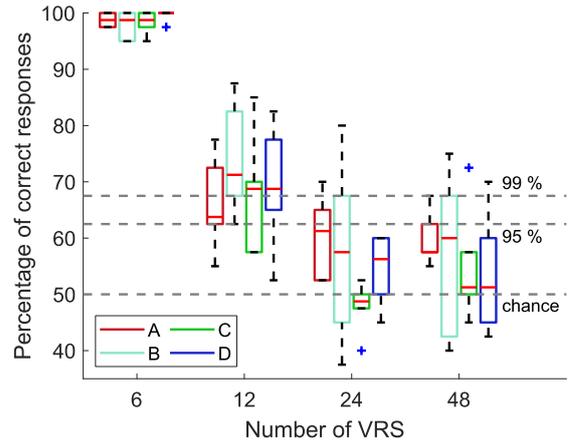

Fig. 5. Percentage of correct responses for different numbers of VRS at position A-D in the corridor. Boxes represent the 25% to 75% percentile range and the horizontal red lines within the boxes represent the median. Whiskers denote highest or lowest value within 1.5 times the interquartile range from the box edge. Any values outside of this range were classified as outliers, represented by a blue cross. The different virtual room conditions are represented by different colors of the boxes. Along the x-axis, the number of VRS is denoted. The three horizontal lines indicate the chance level and the 95% and 99% confidence range of a binomial distribution with $p = 0.5$.

conditions. No significant difference was found between 24 and 48 VRS.

Taken together, the listening test indicates that for the pink pulse stimulus and the corridor tested in the current study, a minimum of 24 VRS appears sufficient to render the late reverberation.

## VI. DISCUSSION

A computationally efficient method to render late reverberation in rooms with inhomogeneously distributed absorption coefficients was proposed, extending the perceptually plausible room acoustics simulator (RAZR, [24]). The technical and perceptual evaluation showed that a relatively low number of 24 spatially equally distributed virtual sources to auralize late reverberation approximates the reference condition with 96 VRS sufficiently well. With relaxed criteria, depending on the application, 12 VRS might also be sufficient. Thus, computational cost for simulating late reverberation and demands for spatially rendering late reverberation can be kept low, which is particularly important in the context of real-time VAEs. Although listening tests were conducted with a specific room acoustics simulation method using an FDN, the current results will apply to different simulations methods for late reverberation, provided a number of incoherent late reverberation signals is generated.

*A. Limitations of the suggested warping-based spatial subsampling method*

The suggested direction-warping-based spatial subsampling method to derive weights for the blending of different absorption coefficients at the room boundaries can be efficiently applied using only vector-based operations for the warping and the well-established VBAP method for blending. Moreover, for an axis-aligned "proxy" shoebox room (orthogonal walls), VBAP simplifies to just using the absolute values of the corresponding elements of the warped VRS vectors (per octant), dramatically reducing computational complexity.

For low numbers of VRS and when the direction of the VRS coincide with the surface normal, which form the vector base for VBAP in each octant, the method becomes invariant to the FOV under which a boundary occurs. This is obvious in the current study for 6 VRS, where, e.g., the IC is independent of position (see Fig. 3, position A and D). The method can be further improved, mainly for low numbers of VRS, by integrating an estimate of the FOV of each boundary and the average FOV covered by each VRS. However, relevant improvements would be mainly observed in position D and for the lowest number of 6 VRS, which cannot be recommended for application in VAEs, anyhow.

Furthermore, the current method is based on a "proxy" shoebox room approximation of the underlying room geometry, which already represents spatially averaged absorption coefficients in six orthogonal directions. Future developments for arbitrary geometry can be based on image down-sampling methods after projection and rendering of the environment to a cube or spherical map, as applied in computer graphics/vision.

*B. Physically-based parameter choices for the FDN*

The current results indicate that a relatively low number of 24 VRS is sufficient to render late reverberation. For high computational efficiency an according low number of FDN channels is desirable. The lower the number of FDN channels, the more pronounced recurrent temporal features occur in the FDN output. To avoid any perceptual difference due to varying the number of FDN channels as such, in the current study, the number of FDN channels was kept fixed at 96, independent of the number of VRS. For real-time applications, a perceptually optimal choice of FDN parameters for a low number of FDN channels has to be found. While several studies optimized FDNs using, e.g., time-varying parameters or optimization of mode density (e.g., [43], [34]), RAZR uses a physically-based design of the FDN, where the delays are derived from the dimensions of the room. This is also motivated by interpreting the FDN as a rough approximation of radiance transfer [44]. The current default settings as suggested and evaluated in [24] use 12 VRS for lowest computational complexity. According to the current finding, an optimal representation of random sound directions travelling through the room has to be found for 24 VRS. This higher number allows for a better sampling of the distribution of sound travelling times between two boundaries as, e.g., calculated from radiance transfer.

*C. Relation of technical and perceptual evaluation and outlook*

The observations of the technical evaluation with accurate representation of the IC up to at least 1.5 kHz for 24 and more VRS and small deviations in the ILD for 24 and more VRS are in line with the absence of any significant differences for 24 and more VRS in the perceptual discrimination experiment. With respect to recommendations for spatial resolution of late reverberation in VAEs, 24 VRS thus appear sufficient both from the technical and the perceptual side.

However, differences to 12 VRS are not large. Moreover, it can be assumed that other source signals like speech or random noise bursts are less critical for the perceptual evaluation than the here employed pulse stimulus, and requirements for late reverberation might be further relaxed depending on the targeted application. The effect of receiver orientation was only assessed in the technical evaluation, however, it appears plausible that the lowest number of 6 VRS will also lead to large perceptual deviations from higher numbers of VRS and can be considered unsuited for in the context of VAEs, where a 6-DOF movement of the receiver is generally desired. Furthermore, for a more complete picture also on the side of the technical evaluation, future research could be extended to separately assess the effects of the number of VRS and the topology of the loudspeaker array, additionally considering isotropic cases. For the use of VAEs in the context of an ecologically valid evaluation of hearing aids or hearing supportive devices, a technical assessment of two receivers spaced at typical multi-microphone distances of ear-level devices appears relevant.

VII. CONCLUSION

A highly efficient method for rendering anisotropic late reverberation for VAEs via a limited number of VRS has been proposed. For a "proxy" shoebox representation of arbitrary room geometry, the (average) absorption coefficients assigned to each boundary are smoothly mapped to a desired number spatially equally distributed VRS, depending on the receiver position in the room. For an axis-aligned proxy shoebox, the required operations can be strongly reduced and computational efficiency further improved. Technical evaluation in terms of IC and ILD suggests that across multiple positions in a room with strongly inhomogeneous absorption properties, a number of 24 VRS can suffice for accurate reproduction in 6-DOF applications. A listening test confirmed this result. Further reduction of the number of VRS may be possible in certain applications.

The room acoustics simulator RAZR [24] including the current extensions is freely available at www.razrengine.com.